\documentclass{PoS}
\usepackage{epsfig}
\PoS{PoS(LAT2005)253}

\title{Cutoff effects in the O(N) sigma model at large N. }

\ShortTitle{Cutoff effects in the O(N) sigma model at large N. }

\author{\speaker{Ulli Wolff} and Francesco Knechtli\\
        HU Berlin, Newtonstr. 15, 12489 Berlin, Germany\\
        E-mail: \email{uwolff@physik.hu-berlin.de, knechtli@physik.hu-berlin.de}}

\author{Bj{\"o}rn Leder\\
        DESY, Platanenallee 6, 15738 Zeuthen, Germany\\
        E-mail: \email{bjoern.leder@desy.de}}
\author{Janos Balog\\
        RMKI, 1525 Budapest 114, Pf.49, Hungary\\
        E-mail: \email{balog@rmki.kfki.hu}}

\abstract{The computation of the step scaling function for the finite size
mass-gap in the O($N$) sigma model at large $N$ is reviewed. Practically exact 
nonperturbative results become available for both
finite and vanishing lattice spacing. We use them as a testbed to investigate
standard procedures of continuum extrapolation in lattice field theory.
\vspace{3cm}
\begin{flushright}
HU-EP-05/47\\
SFB/CPP-05-47
\end{flushright}
}

\FullConference{XXIIIrd International Symposium on Lattice Field Theory\\
		 25-30 July 2005\\
		 Trinity College, Dublin, Ireland}

\newcommand{\be}{\begin{equation}}
\newcommand{\ee}{\end{equation}}

\newcommand{\rO}{\mathrm{ O}}
\newcommand{\mathe}{\mathrm{e}}

\begin{document}

\section{Introduction}
Numerical simulations in lattice field theory are
only possible when both the lattice spacing and the volume
are finite. If the thermodynamic limit is to be taken in theories
with a mass gap, it is approached exponentially fast and usually
the extrapolation causes no essential problems. Often, one is even
interested in the universal continuum limit at a finite volume.
It remains as the main problem to extrapolate the lattice spacing $a$
to zero. Normally the use of some analytic formula for (at least) the asymptotic
dependence of physical results on $a\to 0$ is  unavoidable\footnote{
See however \cite{Hasenbusch:2002}
for an interesting attempt to avoid an explicit such step.}.
Often this is based on the general framework 
given by Symanzik \cite{Symanzik:Cutoff},
which for a universal dimensionless quantity $P$ in
many theories, including the $\sigma$-model, amounts to setting
\be
P(a) \simeq P(0) + \frac{a^2}{L^2} \left[c_0 +c_1 \ln(a/L) + c_2 \ln^2(a/L) 
                                   +\ldots \right] +\rO(a^4),
\label{Symfit}
\ee
where the coefficents $c_i$ have a finite continuum limit. In particular,
in perturbation theory they contain only renormalized couplings,
the bare one is eliminated before organizing the series in powers of ln.
For $L$ we may take any finite renormalized length scale,
an inverse mass for example.
In finite
volume calculations it is convenient to use the system extension.
The derivation is based either on effective field theory ideas or
on perturbation theory. There the power of the logarithm grows with the
loop order. In applications in connection with Monte Carlo simulations
one normally assumes that the function in the square bracket varies slowly
over the range of accessible lattice spacings and it is often replaced by
a constant. Then typically the form $A+B (a/L)^2$ is fitted to the data
after discarding results for larger $a/L$ until the fit is
acceptable in terms of $\chi^2$. 
Then the fit parameter $A$ inherits an
error $\delta A$
from the data by error propagation, and in this way $P(0)$ is estimated\footnote{
Sometimes
more conservative criteria are invoked at this point to exclude
additional coarse data or add systematic errors.}.
Clearly this procedure inflicts some systematic error, which may not
always be negligible in high precision simulations as they are hopefully
becoming more and more common in QCD in the future.
We hence here take the occasion to investigate this issue in a simplified
model where we have access to exact but still nontrivial nonperturbative
results both in the continuum and at finite lattice spacing.

In cluster simulations of the two-dimensional 
O(3) nonlinear $\sigma$-model
\cite{Hasenfratz:2001,Hasenbusch:2002} over a rather 
large range of small $a$ a behavior was 
empirically found for some quantities, 
which seemed hard to describe as $a^2$ but rather looked
like a linear dependence. In \cite{Knechtli:2005jh} we report additional
simulations at $N=4$ and $N=8$ and a leading and subleading
large $N$ calculation. The main conclusion is that the 
often neglected $\ln$ terms
in (\ref{Symfit}) can and in fact do at large $N$ 
mimic a behavior of the kind found in the simulations.
We extend this investigation here by performing various
extrapolations with the large $N$ data and then assessing the systematic
deviation from the known continuum limit.
As a warning
we shall demonstrate cases of significant discrepancy in
combination with acceptable $\chi^2$ and a plausible
extrapolation formula.

\section{Step scaling function of the $\sigma$-model at large $N$}

We start directly from the lattice formulation of the model
where on a two-dimensional cubic lattice of size $T\times L$
the unit length $N$-component spin field $s(x)$ with partition
function
\be
 Z = \int \prod_x d^N s \delta ( s^2 - 1 ) \, \mathe^{- S ( s )}
  \label{Zsigma}
\ee
is governed by the euclidean action
\be
S ( s ) = \frac{N}{2 \gamma} \sum_{x \mu} ( \partial_{\mu} s )^2.
\ee
Here $\partial_{\mu}$ is the forward difference operator and $\gamma$ the rescaled
coupling (held fixed in the large $N$ limit).

As a physical observable that is finite in the asymptotically free
(according to conventional wisdom) continuum limit we study the finite volume
mass-gap \cite{Luscher:1982uv} $M(L)$ that can be extracted 
on a torus with $T\to\infty$
at finite $L$ from the decay of the fundamental spin correlation at zero spatial
momentum. More precisely we study its scale dependence via the step
scaling function (SSF) \cite{stepscaling}
\be
\label{Sigma}
    \Sigma(u,N,a/L) = M(2L)\,2L \quad\mbox{for}\quad M(L)L=u
\ee
under factor-two rescalings of $L$ with the continuum limit
\be
\lim_{a/L\to0}\Sigma(u,N,a/L)=\sigma(u,N).
\ee

The large $N$ evaluation of the path integral (\ref{Zsigma}) is a standard
saddle point calculation.
%The length constraint $\delta ( s^2 - 1 )$ at each lattice site is
%Fourier-represented. Then the action becomes gaussian in the spin field
%which can hence be integrated out. For the effective action for
%the `composite' field consisting of the Fourier variables a saddle
%point expansion is made which is organized in powers of $1/N$
%(analogous to $\hbar$ in quantum theories). 
More details on these steps
can be found in \cite{Knechtli:2005jh}. At leading order gaussian fluctuations of $s(x)$ are
controlled by the (euclidean) lattice Klein Gordon operator $-\Delta + m_0^2$
where the mass parameter
$m_0$ is fixed dynamically in terms of $\gamma$ and $a/L$
by the gap equation for $T=\infty$
\be
\frac{1}{\gamma} = \frac{1}{L} \sum_{p = \frac{2 \pi}{L} n} \frac{1}{2
   \hat{\omega} ( p ) \sqrt{1 + a^2 \hat{\omega}^2 ( p ) / 4}}, \quad
   \hat{\omega}^2 = m_0^2 + \hat{p}^2, \quad \hat{p}=\frac2{a} \, \sin(ap/2).
\label{gapeq}
\ee
The finite volume mass-gap is then closely related,
$ % \be
\sinh(aM/2)=am_0/2\, .
$ %\ee

If we expand the SSF as
\be
\Sigma(u,N,a/L)=\Sigma_0(u,a/L)+\frac1{N} \Sigma_1(u,a/L) + \rO(1/N^2),
\ee
the leading term $\Sigma_0$ is obtained by twice using (\ref{gapeq}) --
for some resolution $L/a$ and the doubled value -- and by combining the
two to eliminate $\gamma$. Evaluated numerically, this yields the exact
SSF at finite cutoff. By an asymptotic expansion in powers of $a$ using results
of \cite{Caracciolo:1998gg} we obtain the continuum value $\sigma_0(u)$
and the form of the leading artefacts.
%The limit
%\be
%\sigma_0(u)=\lim_{a/L \to0}\Sigma_0(u,a/L) 
%\ee
The limit is determined by the transcendental equation
\be
F ( \sigma_0 ) = F ( u ) - \ln 2,
   \quad F ( z ) = \frac{\pi}{z} + \sum_{n = 1}^{\infty}
   \left( \frac{1}{\sqrt{n^2 + (z/2\pi)^2}} - \frac{1}{n} \right) .
\label{sigma0}
\ee
This continuum equation can be expanded in powers of $u$
thus reproducing the large $N$ limit of the known orders
of perturbation theory of the $\beta$-function
for this renormalized coupling
\cite{Luscher:1982uv,stepscaling,Shin:1998bh,Knechtli:2005jh}. 
For large $u$ the mass-gap becomes
insensitive to the volume and $\sigma_0(u)$ tends to the value $2u$
exponentially
fast. 
%ML, LWW, Shin, KLW
The asymptotic lattice correction to $\sigma_0$ is rigorously known to be of the
form (\ref{Symfit}) with only $c_0,c_1$ non-vanishing. 

\begin{figure}
\begin{center}
\epsfig{file=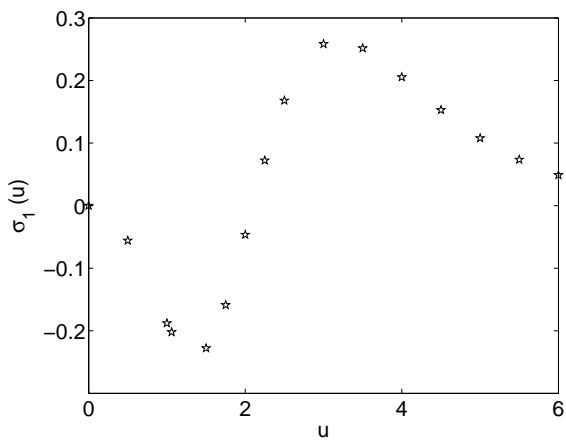, width=0.5\textwidth}
\caption{Correction at order $1/N$ of the continuum step scaling function.}
\label{sigma1}
\end{center}
\end{figure}

To obtain the $1/N$ correction $\Sigma_1$ we have computed the
leading self-energy correction of the spin two-point function that
shifts its pole by an order $1/N$ term. 
%It consists of two simple diagrams
%which contain however the rather complicated propagator of the new
%field related to the constraint.
The diagrams could only be evaluated \cite{Knechtli:2005jh}
numerically including a numerical integration. We managed however to compute
them to many digits and could extract the continuum limit shown in Fig.~\ref{sigma1}.
%
%It is worth noting that this finite result emerges after cancelling a divergence
%between the two diagrams, and that 
The correction vanishes for both large
and small $u$, since $\sigma$ is fixed uniformly in $N$ in
these limits. 

The investigation of the analytic behaviour of this correction is very hard.
We know a representation of the continuum limit in terms of a numerical
integral and a two-fold infinite sum which agrees with the lattice continuum
extrapolation. However, it seems actually easier to perform the latter, as
the sums also need an extrapolation. The small $u$ expansion of the continuum
form could be shown to agree with the $1/N$ terms in the perturbative $\beta$-function
up two three loop order.
An extraction of the form of the asymptotic cutoff
dependence has so far been beyond our capabilites.
There is
however strong numerical evidence \cite{Knechtli:2005jh}
that it is of the same type as in the leading term, i.~e. just
$a^2$ and $a^2 \ln a$.

\section{Sample continuum extrapolations of large $N$ data}

We now use our `exact' data for the nonperturbative values of
$\Sigma_0, \Sigma_1$
to mimic their extrapolation based on lattices that
could be simulated and with superimposed statistical noise.
We test various more or less plausible fit functions.
Our question is: when does this lead to significant biases beyond the
error determined during the extrapolation?

We consider two sequences of lattices for our virtual simulations
\be
{\cal L}_{\rm poor} = \{ 8, 12, 16, 24 \},\quad
{\cal L}_{\rm rich} = \{ 10, 12, 16, 24, 32, 64, 96, 128 \}.
\label{lats}
\ee
For the lattices of either set we add artificial errors to $\Sigma_0$
\be
S_i=\Sigma_0(u,L_i/a) + w \eta_i \, , \quad i=1,\ldots,n,
\ee
where $\{\eta_i \}$
are gaussian random numbers of unit variance and $w$ controls the size 
of the errors. We fit these numbers to a combination of $n_f$
independent fit-functions 
\be
S_i \approx F_i=F(L_i/a)=\sum_{\alpha=1}^{n_f} A_{\alpha} f_{\alpha}(L_i/a),
\ee
where we always take $f_1=1$ such that $A_1$ is the extrapolation. It is now
a simple matter of linear algebra to show that on average (over the $\{\eta_i \}$)
we find
\be
\frac{\chi^2_{\rm min}}{n-n_f} = 1 + \frac1{w^2(n-n_f)}
\sum_{ij} \Sigma_0(u,L_i/a) \bar{P}_{ij}  \Sigma_0(u,L_j/a)
\label{chimin}
\ee
with $\bar{P}$ the projector to the $(n-n_f)$-dimensional
space orthogonal to the one spanned
by the $f_{\alpha}(L_i/a)$. We define that for a given set of
functions a fit is accepted if $\chi^2_{\rm min}/(n-n_f) \le 2$
holds. Formula (\ref{chimin}) allows now to determine a value $w_0$
such that this is true as long as $w \ge w_0$.
By similar algebra we find the mean propagated extrapolation error
\be
\delta A_1=w\sqrt{(M^{-1})_{11}}
,\quad 
%$ with $
M_{\alpha\beta}=
         \sum_i f_{\alpha}(L_i/a) f_{\beta}(L_i/a)
\ee
for the mean value of $A_1$ obtained by solving the linear system
\be
\sum_{\beta} M_{\alpha\beta}A_{\beta}=\sum_i f_{\alpha}(L_i/a)\Sigma_0(u,L_i/a).
\ee

\begin{figure}
\begin{center}
\epsfig{file=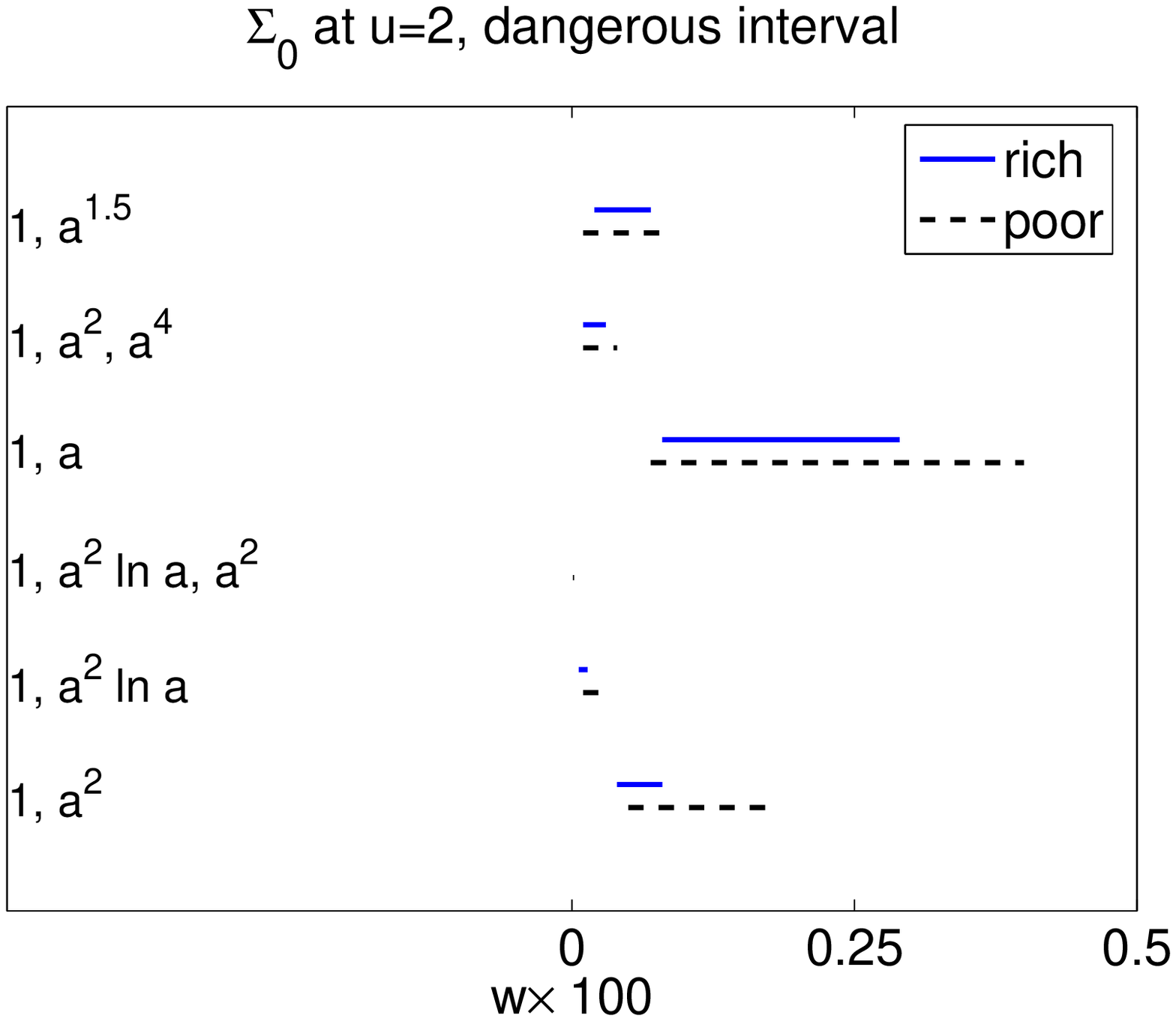, width=0.45\textwidth}
\epsfig{file=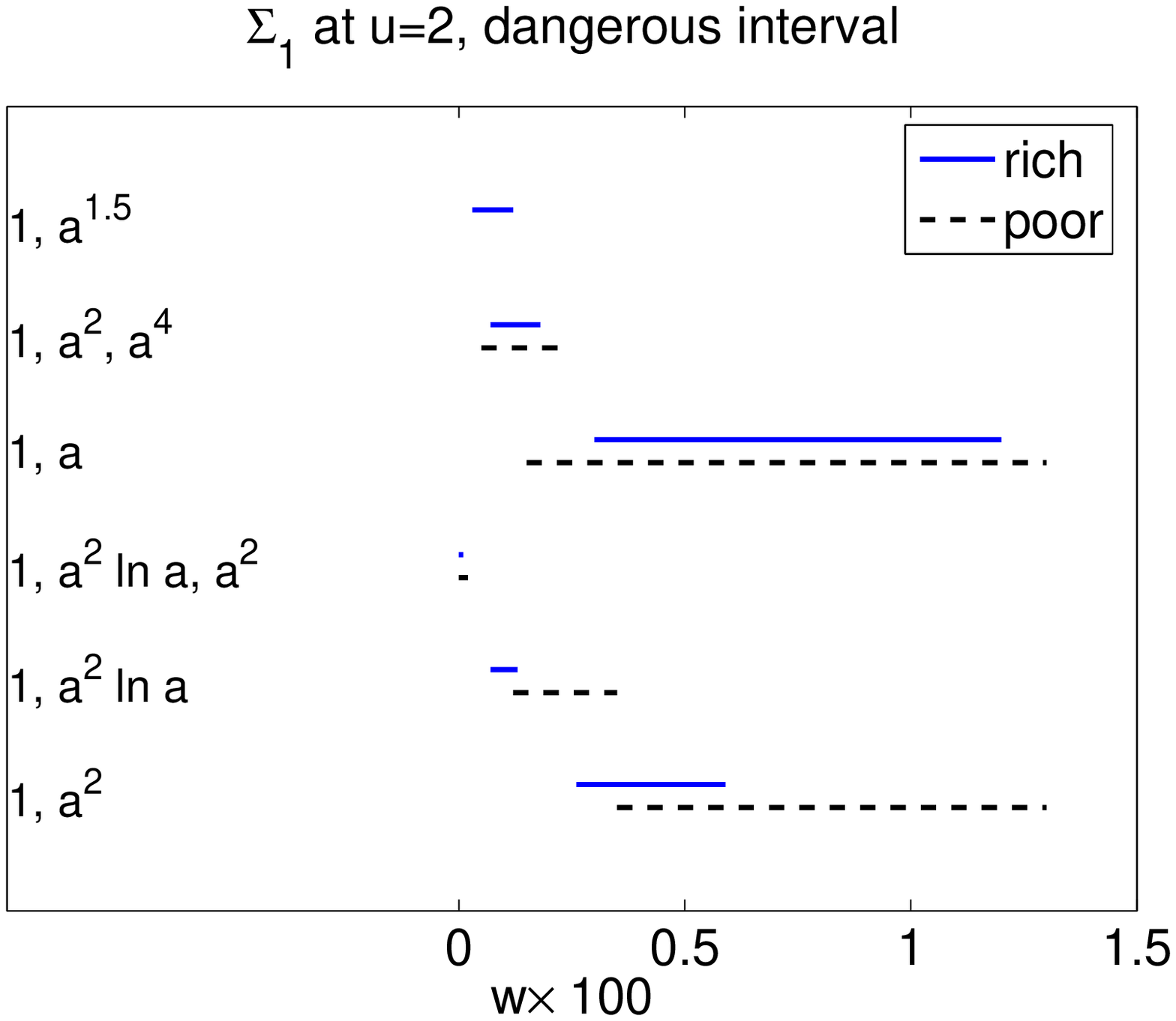, width=0.45\textwidth}
\caption{Dangerous intervals for extrapolations  for $\Sigma_0$ (left)
and $\Sigma_1$ (right).
The fit-functions employed are 
shown in the left of the figures, and solid and dashed lines refer to
the sequences of lattices in (\protect\ref{lats}).}
\label{intervals}
\end{center}
\end{figure}

This in turn implies an error level, such that for $w \ge w_1$
the systematic error is not significant in the sense 
$|A_1-\sigma_0(u)| \le \delta A_1$. In other words, for $w_0 \le w \le w_1$
we typically find extrapolations that look acceptable but are
intolerably biased. Of course, this is excluded, if $w_0 \ge w_1$ holds.
We now use our data for $\Sigma_0$ and $\Sigma_1$ to determine the
`dangerous interval' $[w_0 , w_1]$ for a number of cases. We choose the value
$u=2$ where the cutoff effects, which are in general rather small at large $N$,
are most noticeable for $\Sigma_0$. 
Our results are summarized in Fig.~\ref{intervals}.
The absolute errors
$w$ should be compared compared to 
$\sigma_0 (2)=3.2726 $ and $\sigma_1 (2)= -0.0467 $.

\section{Conclusions}
We have sketched  the computation of the step scaling
function of the O($N$) nonlinear $\sigma$-model in the large $N$ limit
including the subleading order. We focused on the approach to the
continuum limit of these quantities which is reached with
corrections of the form $a^2$ and $a^2 \ln a$. We used these
data known beyond Monte Carlo precision to investigate various
extrapolation methods, for instance in terms of pure powers.
It turned out
that for small enough errors there are cases
possible with acceptable
$\chi^2/\mbox{d.g.f} \le 2$ and an extrapolation bias beyond the statistical
error of the extrapolation. As expected this becomes less likely
if the assumed form is close to the true one and/or if the
range of lattice spacings measured is large (except for the fit to $a^{1.5}$).
As a very minimal precaution, one should try also forms with
logarithms beside pure powers to probe systematic errors.
The whole investigation may also be interpreted as a strong case
for improvement programmes since flat data are easy to extrapolate.

\noindent {\bf Acknowledgements.}  We would like to thank  Peter Weisz
for helpful discussions and him and Rainer Sommer for 
a critical reading of the manuscript.
This work was supported by the Deutsche Forschungsgemeinschaft
in the form of Graduiertenkolleg GK~271 and Sonderforschungsbereich SFB~TR~09.

%\begin{thebibliography}{99}

%\end{thebibliography}
\providecommand{\href}[2]{#2}\begingroup\raggedright\endgroup

%\bibliography{references}
%\bibliographystyle{JHEP}
\end{document}